\documentclass[letterpaper]{article} 
\usepackage{aaai24}  
\usepackage{times}  
\usepackage{amsmath}
\usepackage{amssymb}
\usepackage{tabularx}
\usepackage{rotating}
\usepackage{helvet}  
\usepackage{courier}  
\usepackage[hyphens]{url}  
\usepackage{graphicx} 
\usepackage{natbib}  
\usepackage{caption} 
\usepackage{xcolor}
\usepackage{lipsum}
\usepackage{makecell}
\usepackage{algorithm}
\usepackage{algorithmic}

\usepackage{newfloat}
\usepackage{listings}
\DeclareCaptionStyle{ruled}{labelfont=normalfont,labelsep=colon,strut=off} 
\floatstyle{ruled}
\newfloat{listing}{tb}{lst}{}
\floatname{listing}{Listing}
\title{International Trade Flow Prediction with Bilateral Trade Provisions}
\author{
    Zijie Pan, Stepan Gordeev, Jiahui Zhao, Ziyi Meng, Caiwen Ding, Sandro Steinbach, Dongjin Song\\
}
\affiliations{
    Department of Computer Science and Engineering, University of Connecticut, Storrs, CT, USA\\

    \{zijie.pan, stepan.gordeev, jiahui.zhao, nicole.meng, caiwen.ding, sandro.steinbach, dongjin.song\}@uconn.edu
}

\usepackage{bibentry}

\begin{document}

\maketitle

\begin{abstract}

This paper presents a novel methodology for predicting international bilateral trade flows, emphasizing the growing importance of Preferential Trade Agreements (PTAs) in the global trade landscape. Acknowledging the limitations of traditional models like the Gravity Model of Trade, this study introduces a two-stage approach combining explainable machine learning and factorization models. The first stage employs SHAP Explainer for effective variable selection, identifying key provisions in PTAs, while the second stage utilizes Factorization Machine models to analyze the pairwise interaction effects of these provisions on trade flows. By analyzing comprehensive datasets, the paper demonstrates the efficacy of this approach. The findings not only enhance the predictive accuracy of trade flow models but also offer deeper insights into the complex dynamics of international trade, influenced by specific bilateral trade provisions.

\end{abstract}

\section{Introduction}

International bilateral trade flow serves as an important economic indicator, which represents the value of goods and services that have been exported from one country
to another. It is used by economists and policymakers and influences international trade policy as well as domestic economic policy in both countries.  For example, an increase in exports from China to Cambodia  would exacerbate Cambodia’s trade balance (i.e. exports minus imports.) As a result, Cambodia may need to fill the financial shortfall created
by this increased outflow of money to China, while China will likely benefit from the influx of foreign assets from Cambodia. Studying bilateral trade flow becomes an important research topic since it is a major determining factor in the process of a country’s economic development. On the other hand, due to the challenges encountered in multilateral trade discussions within the World Trade Organization (WTO) over the past twenty years, nations have progressively shifted their attention to preferential trade agreements (PTAs) that encompass just one or a few select partners. Existing research has tried to model the overall impact of PTAs and to establish the relative importance of individual trade agreement provisions in the agreement's overall impact~\cite{kohl2016trade,dhingra2018beyond,regmi2020using, falvey2022breadth,mattoo2022trade}. However, these attempts face the challenges of a large number of provisions and multicollinearity issues, which make it very difficult to analyze the individual and pair-wise effects of provisions. In this paper, to investigate the effects of provisions in international bilateral trade, we propose a two-stage approach, that first leverages explainable machine learning for variable selection and second models the pair-wise interaction effects among provisions using state-of-art factorization models



\section{Related Work and Contribution}

Existing economic literature that studies bilateral trade flow majorly adopts an empirical method called the Gravity Model of Trade, which is motivated by Newton’s law of universal gravitation. Tinbergen \cite{tinbergen1962shaping} proposed to model the bilateral trade flows between two countries using GDPs of the origin and destination countries as well as the distance between the two countries:

\begin{equation}
\text{FLOW}=\alpha \frac{\text{GDP}_o^{\beta_1} \text{GDP}_d^{\beta_2}}{\text{DIST}^{\beta_3}}
\end{equation}

Ever since the model has been used extensively because of its empirical power. To identify provisions that promote agricultural trade, a three-way multiplicative gravity model is also widely used \cite{yotov2016advanced}, which represents the expected trade flows as an exponential function of relevant covariates, along with three sets of high-dimensional fixed effects.  
These fixed effects account for multilateral trade resistances and unobserved time-invariant trade costs\cite{weidner2021bias}. Within the three-way gravity
framework, we tackle the empirical challenges of dealing with numerous correlated covariates and an abundance of zero observations simultaneously. Starting with an empirical model that assesses the relationship between trade flows $X_{ijt}$ and provisions $\tau_{ijt}$:

\begin{equation}
\begin{aligned}
\mu_{i j t} &:= \mathrm{E}\left(X_{i j t} \mid \tau_{i j t}^{\prime}, \alpha_{i t}, \gamma_{j t}, \delta_{i j}\right) \\
&= \exp \left(\tau_{i j t}^{\prime} \beta^{\prime} + \alpha_{i t} + \gamma_{j t} + \delta_{i j}\right)
\end{aligned}
\end{equation}

where $i$, $j$, and $t$ denote the exporter, importer, and year respectively. $X_{ijt}$ indicates agricultural exports from country $i$ to country $j$ in year $t$ and $\tau_{i j t}^{\prime}$ denotes the vector of provisions, which includes each provision in an enforced bilateral or regional trade agreement. It also accounts for the multilateral trade resistances with the high-dimensional fixed effects $\alpha_{i t}$ and $\gamma_{j t}$. In addition, it includes time-invariant exporter-importer fixed effects $\delta_{i j}$, which account for unobserved trade costs potentially correlated with the provisions.
Since Equation 2 fails to identify the parameters of interest accurately due to overfitting and multicollinearity. Such a challenge arises due to the large number of provisions likely correlated with each other and only a subset of those provisions will have a binding (non-zero) effect on trade flow. \cite{breinlich2022machine} relies on a plug-in Lasso regularized regression approach, which specifies the regression model to be consistent with the gravity model of international trade. Formally, it amends the minimization problem that defines the three-way gravity by adding a penalization term that purges provisions with coefficients equal to zero:

\begin{equation}
\begin{aligned}
(\hat{\alpha}, \hat{\gamma}, \hat{\delta}, \hat{\beta}) &:= \arg \min_{\alpha, \gamma, \delta, \beta} \frac{1}{n}\left(\sum_{i, j, t}\left(\mu_{i j t} - X_{i j t} \ln \mu_{i j t}\right)\right) \\
&\quad + \frac{1}{n} \sum_{l=1}^m \lambda \hat{\phi}_l\left|\beta_l\right|
\end{aligned}
\end{equation}

following the same notation in Equation 3, $n$ is used to indicate the number of observations. The first part of Equation 3 represents the standard Poisson Pseudo Maximum Likelihood (PML) minimization problem using the pseudo-likelihood function, while the second part is the Lasso penalty term, which consists of two tuning parameters, $\gamma \leq 0 $ and $\hat{\phi}_l \geq 0$. Refining the model iteratively across provisions, the tuning parameters shrink the $\beta$ coefficient to zero.
None of the aforementioned methods take variable importance into variable selection account and ignore the coexistence effects of provisions during trade flow prediction. To amend the shortcomings of existing research, we contribute to providing a more explainable approach to select prominent provisions for further analysis and explicitly model the pair-wise effects of provisions for trade flow predictions.

\section{Data}

This analysis includes data on international trade flows from UN Comtrad, covering all agricultural trade flows between 1968 and 2017. We combined the trade flow data with the Deep Trade Agreements (DTAs) database, which was collected by Mattoo et al \cite{mattoo2020handbook}, for the content and evolution of bilateral and regional trade agreements. The dataset on the content of trade agreements includes information on 282 PTAs that have been signed and notified to the WTO between 1958 and 2017. Specifically, we treated country pairs without export information as zeros and eliminated country pairs with no provision agreements available.

\section{Methodology}

\subsection{SHAP Explainer}

In game theory, there is a task to "fairly" assign each player a payoff from the total gain generated by a coalition of all players. Formally, let $N$ be a set of $n$ players, $v: 2^N \rightarrow \mathbb{R}$ is a characteristic function, which can be interpreted as the total gain of the coalition $N$. Starting from $v(\emptyset)=0$. Given a coalitional game, the Shapley value \cite{shapley1953value} is a solution to the payoff assignment problem. The payoff (attribution) for player $i$ can be computed as follows:

\begin{equation}
\phi_i(v)=\frac{1}{|N|} \sum_{S \subseteq N \backslash\{i\}}\left(\begin{array}{c}
|N|-1 \\
|S|
\end{array}\right)^{-1}(v(S \cup\{i\})-v(S))
\end{equation}

where $v(S \cup\{i\})-v(S)$ is the marginal contribution of player $i$ to the coalition $S$. Another alternative form of the Shapley value is:

\begin{equation}
\phi_i(v)=\frac{1}{|N| !} \sum_{O \in \mathfrak{S}(N)}\left[v\left(P_i^O \cup\{i\}\right)-v\left(P_i^O\right)\right]
\end{equation}
where $\mathfrak{S}(N)$ is the set of all ordered permutations of $N$, and $P_i^O$ is the set of players in $N$ which are predecessors of player
$i$ in the permutation $O$. To adopt the Shapley value to explain arbitrary models, we denote the model as $f$ and let $N$ be all the input features (attributes) and $S$ is a feature subset of
interest $(S \subseteq N)$.  For an input $x$, the characteristic function $v(S)$ is the difference between the expected model output when we know all the features in $S$, and the expected output when no
feature value is known (i.e. the expected output over all possible input), denoted by

\begin{equation}
v(S)=\frac{1}{\mid \mathcal{X}^{N \backslash S \mid}} \sum_{\boldsymbol{y} \in \mathcal{X}^{N \backslash S}} f(\tau(\boldsymbol{x}, \boldsymbol{y}, S))-\frac{1}{\left|\mathcal{X}^N\right|} \sum_{\boldsymbol{z} \in \mathcal{X}^N} f(\boldsymbol{z})
\end{equation}

where $\mathcal{X}^{N}$ and $\mathcal{X}^{N \backslash S}$ are respectively the input space containing
feature sets $N$ and $N\backslash S$. $\tau(\boldsymbol{x}, \boldsymbol{y}, S)$ is the vector composed by $\boldsymbol{x}$ and $\boldsymbol{y}$ according to whether the feature is in $S$.

\subsection{Factorization Machine}

Factorization machines (FM), proposed by \cite{rendle2010factorization}, is a supervised algorithm that can be used for classification, and regression. The strength of factorization machines over the linear regression and matrix factorization includes modeling $n$-way variable interactions, where n is the number of polynomial order, which is usually set to two and an existing fast optimization algorithm associated with factorization machines can reduce the polynomial computation time to linear complexity, making it extremely efficient especially for high dimensional sparse inputs, which is suitable for provision agreement signing. 
Let $x \in \mathbb{R}^d$ denote the feature vectors of one sample, $y$ be the real-valued label. Formally, a 2-way factorization machine is defined as 

\begin{equation}
\hat{y}(x)=\mathbf{w}_0+\sum_{i=1}^d \mathbf{w}_i x_i+\sum_{i=1}^d \sum_{j=i+1}^d\left\langle\mathbf{v}_i, \mathbf{v}_j\right\rangle x_i x_j
\end{equation}

where $\mathbf{w}_0 \in \mathbb{R}$ is the global bias; $\mathbf{w} \in \mathbb{R}$ denotes the weights of the $i$-th variable; $\mathbf{V} \in \mathbb{R}^{d \times k}$ represents the feature embedding; $\mathbf{v_i}$ represents the $i$-th row of $\mathbf{V}$; $k$ is the dimensionality of latent factors. We use $\langle\cdot,\cdot \rangle$ to represent the dot product of two vectors so that $\mathbf{v_i},\mathbf{v_j}$ can effectively model the interaction between the $i^{th}$ and $j^{th}$ feature 

Optimizing the factorization machines in a straightforward method leads to a complexity of $\mathcal{O}\left(k d^2\right)$ since all pairwise interactions require to be computed. To tackle this inefficiency problem, we can reorganize the third term of FM which could greatly reduce the computation cost, leading to a linear time complexity $\mathcal{O}\left(k d\right)$ The reformulation of the pairwise interaction term is as follows:

\begin{equation}
\begin{aligned}
& \sum_{i=1}^d \sum_{j=i+1}^d\left\langle\mathbf{v}_i, \mathbf{v}_j\right\rangle x_i x_j \\
& =\frac{1}{2} \sum_{i=1}^d \sum_{j=1}^d\left\langle\mathbf{v}_i, \mathbf{v}_j\right\rangle x_i x_j-\frac{1}{2} \sum_{i=1}^d\left\langle\mathbf{v}_i, \mathbf{v}_i\right\rangle x_i x_i \\
& =\frac{1}{2}\left(\sum_{i=1}^d \sum_{j=1}^d \sum_{l=1}^k \mathbf{v}_{i, l} \mathbf{v}_{j, l} x_i x_j-\sum_{i=1}^d \sum_{l=1}^k \mathbf{v}_{i, l} \mathbf{v}_{i, l} x_i x_i\right) \\
& =\frac{1}{2} \sum_{l=1}^k\left(\left(\sum_{i=1}^d \mathbf{v}_{i, l} x_i\right)\left(\sum_{j=1}^d \mathbf{v}_{j, l} x_j\right)-\sum_{i=1}^d \mathbf{v}_{i, l}^2 x_i^2\right) \\
& =\frac{1}{2} \sum_{l=1}^k\left(\left(\sum_{i=1}^d \mathbf{v}_{i, l} x_i\right)^2-\sum_{i=1}^d \mathbf{v}_{i, l}^2 x_i^2\right)
\end{aligned}
\end{equation}

\subsection{Two-Stage Analysis}

In this comprehensive study, our aim is to present a detailed, two-phased methodology for forecasting the patterns of international trade flows, with a specific focus on utilizing the entirety of preferential trade agreements.

During the initial phase of our approach, we commence by converting the continuous variables representing trade flows into a binary format, differentiating between the presence (nonzero) and absence (zero) of trade flows. Following this transformation, we employ a neural network to train our model based on this newly formatted dataset. To further enhance our analysis, we integrate the SHAP (SHapley Additive exPlanations) explainer, as introduced by Lundberg et al. \cite{lundberg2017unified}. The implementation of SHAP in our study is crucial, as it aids in pinpointing the preferential trade agreements that play a significant role in the formation of international trade connections.

Progressing to the second stage of our approach, we shift our focus exclusively to the trade flows that are nonzero, indicating an active trade relationship. Additionally, we incorporate information pertaining to the top-k provisions, identified as being of paramount importance in the previous phase. With these elements in hand, we proceed to apply a Factorization Machine regressor, a sophisticated tool that enables us to carry out predictions related to trade flow magnitudes. Moreover, this regressor presents us with the unique opportunity to delve into the pair-wise interactions between different provisions, allowing for a more nuanced understanding of how these agreements collectively influence international trade patterns.

By adopting this two-stage approach, our analysis stands to gain a more accurate and insightful perspective on international trade flows, benefiting from the detailed examination of preferential trade agreements and their intricate interactions. This methodology not only facilitates precise predictions but also lays the groundwork for a deeper exploration of the dynamics governing global trade.

\section{Results}

\subsection{Identification of Significant Provisions}

\begin{figure}
 \centering
  \includegraphics[width=0.5\textwidth]{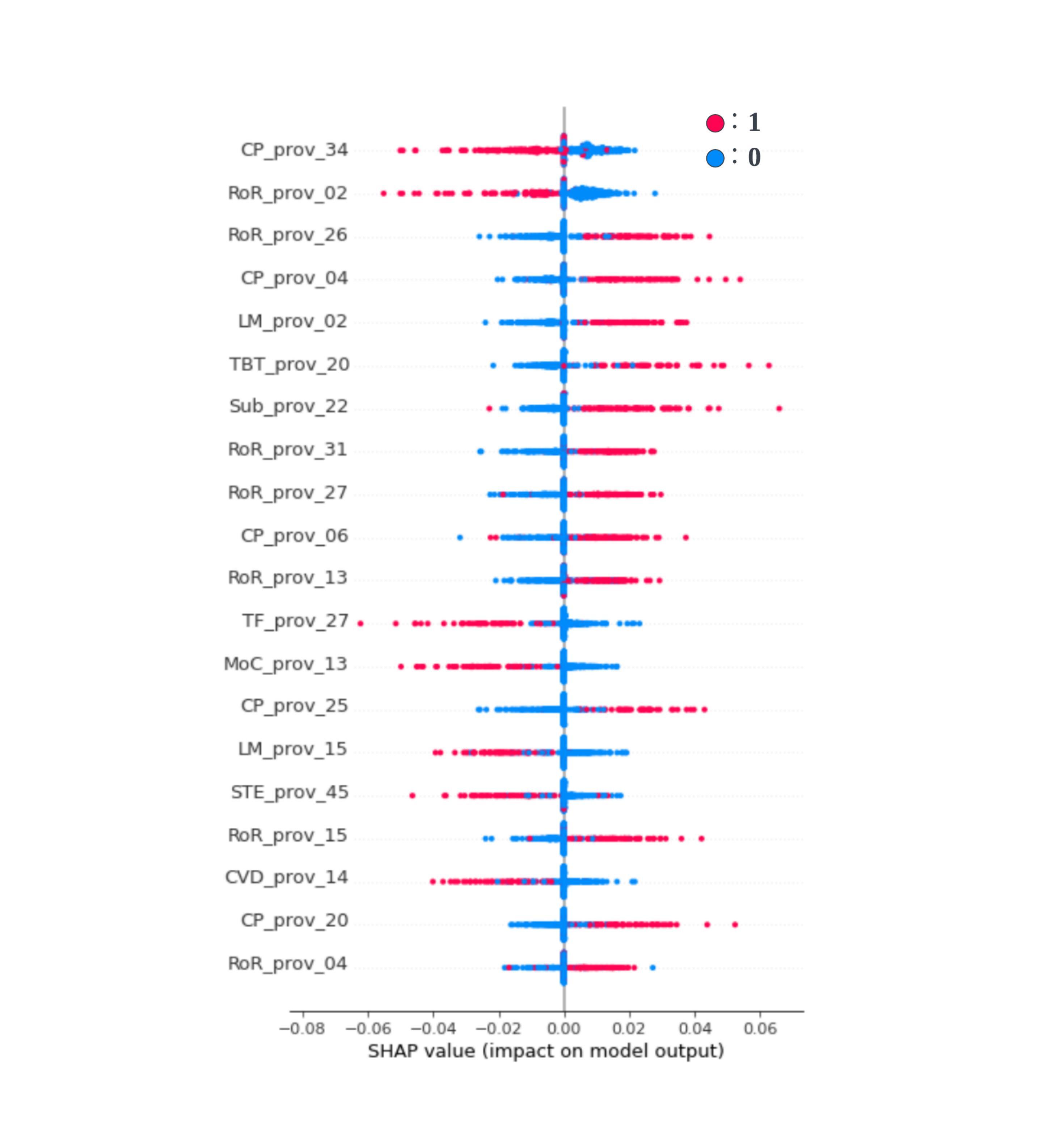}
  \caption{PTA Summary Plot}
  \label{fig:1}
\end{figure}

We adopt a Multi-layer Perceptron (MLP) classifier, which is a kind of artificial neural network to classify the presence of international trade flows based on binary PTA input features. Approximately $96.09 \%$ of the train data was correctly classified by the model, suggesting a high fit to the training data. On unseen test data, the model achieved an accuracy of about $88.29\%$, demonstrating that it generalizes well to new data. We also reported the model's F1 score on the test data is approximately $0.926$. The MLP classifier not only fits the training data well but also generalizes effectively to unseen data, as evidenced by the high test accuracy and F1 score and it is confident in obtaining the significance of provisions using the trained model. We report the top 20 provisions with the largest Shapley values in Table \ref{tab:1} and the corresponding summary plot in  Figure \ref{fig:1}.

\begin{table*}[t]
\scriptsize 
\centering
\caption{Description of Selected  Provisions and Shapley Value}
\begin{tabularx}{\textwidth}{|c|>{\centering\arraybackslash}X|c|}
\hline
\textbf{Provision Id} & \textbf{Description} & \textbf{Values ($\times 10^{-3}$)} \\
\hline
CP 34  & Does the agreement allow for security exceptions?  & 9.07 \\
\hline
RoR 02 & \makecell{Does the certificate have to be issued by competent authorities of the exporting party, \\including customs administrations, other government authorities, and designated private ones?}  & 8.04 \\
\hline
RoR 26 & Is the price basis for the content threshold requirement the FOB (free on board) price?  & 6.60 \\
\hline
CP 04  & Does the agreement promote the principle of transparency? & 6.27 \\
\hline
LM 02  & Does the agreement specify an objective of creation of employment opportunities? & 6.12 \\
\hline
TBT 20 & Is there a regional dispute settlement body? & 6.08 \\
\hline
Sub 22 & Does the agreement provide for any institution to deal with transparency or enforcement? & 6.07 \\
\hline
RoR 31 & Does the agreement contain product-specific rules of origin? & 5.98 \\
\hline
RoR 27 & Is the value content requirement calculated through import content?  & 5.62 \\
\hline
CP 06 & Does the agreement promote the principle of procedural fairness? & 5.23 \\
\hline
RoR 13 & Does the agreement allow for diagonal cumulation?  & 4.93 \\
\hline
TF 27 & Technical assistance and capacity building & 4.92 \\
\hline
MoC 13 & Does the transfer provision apply specifically to transfers relating to repatriation of the proceeds of investment (capital accounts)? & 4.86 \\
\hline
CP 25 & Does the agreement regulates unfair commercial practices? & 4.40 \\
\hline
LM 15 & Does the agreement include reference to cooperation over labor provisions? & 4.39 \\
\hline
STE 45 & Does the agreement provide for any dispute settlement mechanism to deal with state enterprises? & 4.30\\
\hline
RoR 15 & What is the de minimis percentage?  & 4.21 \\
\hline
CVD 14 & Weak CVD Rules (1=weak, 2=real rules) & 3.92 \\
\hline
CP 20 & Does the agreement regulates state aid? & 3.87 \\
\hline
RoR 04 & Is there a certificate exemption?  & 3.82 \\
\hline

\end{tabularx}
\label{tab:1}

\end{table*}

\subsection{Trade Flow Prediction and Provision Interactions}

\begin{figure}
 \centering
  \includegraphics[width=0.5\textwidth]{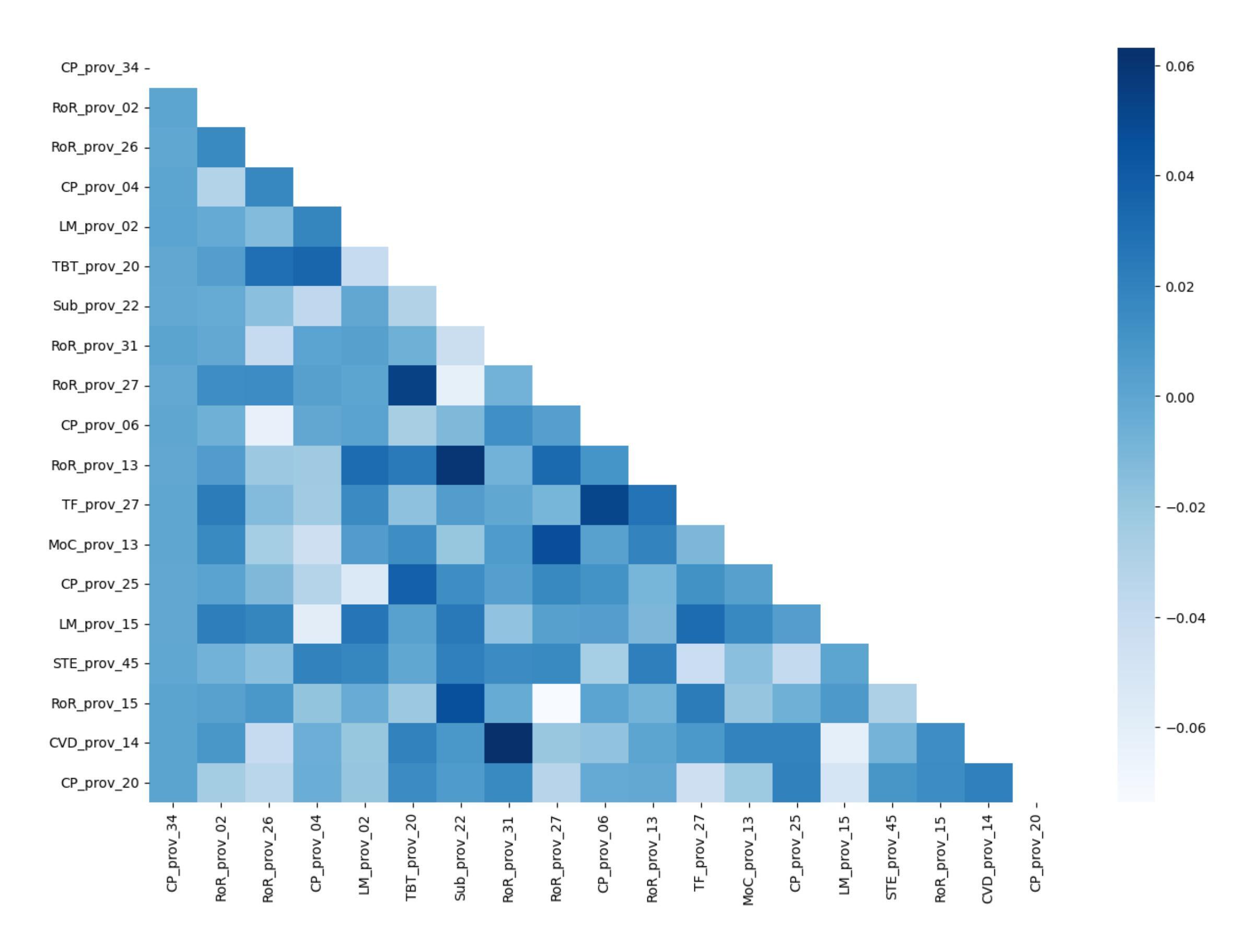}
  \caption{Provision Interaction Plot}
  \label{fig:2}
\end{figure}

 In econometrics and statistics, fixed effects represent the observed quantities that do not vary over the given dataset or are constant across entities. We include the identified important 20 provisions, import country, export country and time index as our covariates so that the three-way fixed effects can be taken into consideration. We aim to use the aforementioned variables to predict the logarithmic values of the trade flows. The regression RMSE of FM is $3.26$ and we display the PTA interaction heat plot as shown in Figure \ref{fig:2}.
The cell value in the heat plot indicates how much the trade flow will increase in log scale if two provisions coexist.

\section{Conclusion}

Existing machine learning methods focus on data security and efficiency perspectives~\cite{xie2023accel, zhao2020connecting, peng2023rrnet, zhao2020knowledge,  luo2023aq2pnn, peng2023autorep, jin2023visual, thorat2023advanced,  jin2024apeer, peng2022length, jiang2024empowering, zhao2022connecting, peng2024maxk, li2023national, jin2024learning, he2024hierarchical, Deng_2024_CVPR, deng2023long, li2023segment, li2024cpseg, shen2024deep, liu2024td3, ni24timeseries}. This research reveals a nuanced understanding of international trade flow influenced by bilateral trade provisions. Key findings demonstrate that specific trade agreements significantly impact trade patterns, underscoring their importance in shaping global economic interactions. This study offers a novel two-stage approach framework by incorporating explainable variable selection method and leveraging advanced regression method that considers pair-wise interactions. It emphasizes the need for detailed analysis of trade provisions to accurately predict and understand trade dynamics, providing a foundation for future research in this field.

\section*{Acknowledgement}
This work was in part supported by the USDA-NIFA Agriculture and Food Research Initiative Program (Award No.: 2022-67023-36399).

\bibliography{aaai24}
\end{document}